\documentstyle[preprint,eqsecnum,prd,aps,epsfig]{revtex}
 \begin{document}

\begin{center}
{\Large 
\bf Confidence intervals for the parameter of Poisson distribution 
    in presence of background\\}
\end{center}

\bigskip

\begin{center}
{\large S.I.~Bityukov$^{a,}$\footnote{\small Corresponding author\\
{\it Email addresses:} bityukov@mx.ihep.su, Serguei.Bitioukov@cern.ch}  
N.V.~Krasnikov$^b$}\\

$^a${\it {\footnotesize Division of Experimental Physics,
Institute for High Energy Physics, Protvino, Moscow Region, Russia}}\\
$^b${\it {\footnotesize Division of Quantum Field Theory,
Institute for Nuclear Research RAS, Moscow, Russia}}\\

\end{center}


\begin{flushleft}
{\large \bf Abstract}\\

\bigskip

A results of numerical procedure for construction of confidence intervals
for parameter of Poisson distribution for signal in the presence of
background which has Poisson distribution with known value of parameter
are presented. It is shown that the described procedure has both Bayesian 
and frequentist interpretations.
\bigskip

{\it Keywords:}  statistics, confidence intervals, Poisson distribution,
Gamma distribution, sample.
\end{flushleft}


\section{Introduction}

In paper~\cite{1} the unified approach to the construction
of confidence intervals and confidence limits
for a signal with  a background presence, 
in particular for Poisson distributions, 
has been proposed. The method is widely
used for the presentation of physical results~\cite{2}
though a number of investigators criticize this approach~\cite{3}

In present paper we use a simple method for construction of
confidence intervals for parameter of Poisson distribution for signal 
in the presence of background which has Poisson distribution with known 
value of parameter. This method is based on the statement~\cite{4} 
that the true value of parameter of the Poisson distribution in the case of 
observed number of events $\hat x$ has a Gamma distribution. 
In contrast to the approach proposed in~\cite{1}, the width of 
confidence intervals in the case of $\hat x = 0$ is independent 
on the value of the parameter of the background distribution.
The described procedure has both Bayesian and frequentist interpretations.
 
In Section 2 the method of construction of confidence intervals 
for parameter of Poisson distribution for signal in the presence of background 
which has Poisson distribution with known value of parameter is described.
The results of confidence intervals construction and their
comparison with the results of unified approach are also given 
in the Section 2. The main results of this note are formulated 
in the Conclusion.

\section{The method of construction of confidence intervals}

Assume that in the experiment with the fixed integral luminosity
(i.e. a process under study may be considered as a homogeneous process
during given time) the $\hat x$ events of some Poisson process
were observed. It means that we have an experimental estimation
$\hat \lambda(\hat x)$ of the parameter $\lambda$ of Poisson distribution.
We have to construct a confidence interval
$(\hat \lambda_1(\hat x), \hat \lambda_2(\hat x))$, covering
the true value of the parameter $\lambda$ of the distribution under
study with confidence level $1 - \alpha$, where $\alpha$ is a
significance level. It is known from the theory of statistics~\cite{5},
that the mean value of a sample of data is an unbiased estimation
of mean of distribution under study. In our case the sample consists
of one observation $\hat x$. For the discrete Poisson distribution
the mean coincides with the estimation of parameter value, 
i.e. $\hat \lambda = \hat x$ in our case. As it is shown in ref~\cite{4}
the true value of parameter $\lambda$ has Gamma distribution 
$\Gamma_{1, \hat x + 1}$, 
where the scale parameter is equal to 1
and the shape parameter is equal to $\hat x + 1$ (see Fig.1),
i.e. 

\begin{equation}
P(\lambda|\hat x) = P(\hat x|\lambda) = 
\displaystyle \frac{\lambda^{\hat x}}{\hat x!} e^{-\lambda}. 
\end{equation}

Note that formula (2.1) results from the Bayesian formula~\cite{6}

\begin{equation}
P(\lambda|\hat x) P(\hat x) = P(\hat x|\lambda) P(\lambda) 
\end{equation}

\noindent
in the assumption that all possible values of parameter $\lambda$
have equal probability, i.e. $P(\lambda) = const$.  
In this assumption the probability that unknown parameter $\lambda$
obeys the inequalities $\lambda_1~\le~\lambda~\le~\lambda_2$ 
is given by evident Bayesian formula 

\begin{equation}
P(\lambda_1 \le \lambda \le \lambda_2|\hat x) = 
P(\lambda_1 \le \lambda|\hat x) - P(\lambda_2 \le \lambda|\hat x) = 
\int_{\lambda_1}^{\lambda_2}{P(\lambda|\hat x)d\lambda}, 
\end{equation}

\begin{center}
$P(\lambda_1 \le \lambda|\hat x) = 
\int_{\lambda_1}^{\infty}{P(\lambda|\hat x)d\lambda}$,
\end{center}

\noindent
where $P(\lambda|\hat x)$ is determined by formula (2.1).

Formula (2.3) has also well defined frequentist meaning.
Using the identity

\begin{equation}
\sum_{i = \hat x + 1}^{\infty}{\frac{\lambda_1^ie^{-\lambda_1}}{i!}} + 
\int_{\lambda_1}^{\lambda_2}
{\frac{\lambda^{\hat x}e^{-\lambda}}{\hat x!}d\lambda} 
+ \sum_{i = 0}^{\hat x}{\frac{\lambda_2^ie^{-\lambda_2}}{i!}} = 1 
\end{equation}

\noindent
one can rewrite formula (2.3) as

\begin{equation}
P(\lambda_1~\le~\lambda~\le~\lambda_2|\hat x) = 
1 - P(n \le \hat x|\lambda_2) - P(n > \hat x|\lambda_1) = 
P(n \le \hat x|\lambda_1) - P(n \le \hat x|\lambda_2), 
\end{equation}

\noindent
where $P(n \le \hat x|\lambda) = \displaystyle
\sum_{n = 0}^{\hat x}{\frac{\lambda^ne^{-\lambda}}{n!}}$ and 
$P(n > \hat x|\lambda) = \displaystyle
\sum_{n = \hat x + 1}^{\infty}{\frac{\lambda^ne^{-\lambda}}{n!}}$.

The right hand side of formula (2.5) has well defined frequentist
meaning  and in fact it is one of the possible definitions of the
confidence interval in frequentist approach~\footnote{See, however,
ref.~\cite{7}}. As an example, such type the shortest 90\% CL confidence 
interval in case of observed number of events $\hat x=4$ is shown in Fig.2.

For instance, for
the case $\lambda_2 = \infty$ formula (2.5) takes the form

\begin{equation}
\int_{\lambda_1}^{\infty}{P(\lambda|\hat x) d\lambda} =
P(\lambda_1 \le \lambda|\hat x) = 
\sum_{n = 0}^{\hat x}{P(n|\lambda_1)} = P(n \le \hat x|\lambda_1), 
\end{equation}

\noindent
which has evident frequentist meaning too. 
 
Let us consider the Poisson distribution with two components:
signal component with a parameter $\lambda_s$ and background component
with a parameter $\lambda_b$, where $\lambda_b$ is known.
To construct confidence intervals for parameter $\lambda_s$ of signal 
in the case of observed value $\hat x$ we must find the distribution 
$P(\lambda_s|\hat x)$. 

At first let us consider the simplest case $\hat x = \hat s + \hat b = 1$.
Here $\hat s$ is a number of signal events and $\hat b$ is a number of
background events among observed $\hat x$ events.

The $\hat b$ can be equal to 0 and to 1.
We know that the $\hat b$ is equal to 0 with probability 

\begin{equation}
p_0 = P(\hat b = 0) = 
\displaystyle \frac{\lambda_b^0}{0!} e^{-\lambda_b} = e^{-\lambda_b} 
\end{equation}

and the $\hat b$ is equal to 1 with probability 

\begin{equation}
p_1 = P(\hat b = 1) = 
\displaystyle \frac{\lambda_b^1}{1!} e^{-\lambda_b} = 
\lambda_b e^{-\lambda_b}.
\end{equation}

Correspondingly, 
$P(\hat b = 0|\hat x = 1) = P(\hat s = 1|\hat x = 1) =
\displaystyle \frac{p_0}{p_0 + p_1}$ and
$P(\hat b = 1|\hat x = 1) = P(\hat s = 0|\hat x = 1) =
\displaystyle \frac{p_1}{p_0 + p_1}$.

It means that distribution of $P(\lambda_s|\hat x = 1)$ 
is equal to sum of distributions 

\begin{equation}
P(\hat s = 1|\hat x = 1) \Gamma_{1,2} + 
P(\hat s = 0|\hat x = 1) \Gamma_{1,1}
= \displaystyle \frac{p_0}{p_0 + p_1} \Gamma_{1,2} +
\displaystyle \frac{p_1}{p_0 + p_1} \Gamma_{1,1},
\end{equation}

\noindent
where $\Gamma_{1,1}$ is Gamma distribution with probability
density $P(\lambda_s|\hat s = 0) = \displaystyle e^{-\lambda_s}$ 
and $\Gamma_{1,2}$ is Gamma distribution with probability
density $P(\lambda_s|\hat s = 1) = \displaystyle \lambda_s e^{-\lambda_s}$.
As a result we have 

\begin{equation}
P(\lambda_s|\hat x = 1) = 
\displaystyle \frac{\lambda_s + \lambda_b}{1 + \lambda_b} 
\displaystyle e^{-\lambda_s}.
\end{equation}

Using formula (2.10) for $P(\lambda_s|\hat x = 1)$ and formula (2.5)
we construct the shortest confidence interval
of any confidence level in a trivial way.

In this manner we can construct the distribution of $P(\lambda_s|\hat x)$
for any values of $\hat x$ and $\lambda_b$. As a result we have obtained 
the known formula~\footnote{The formula (2.11) has been derived
earlier in ref.~\cite{8} (formula 5.88) in the framework of 
Bayesian approach and in ref.~\cite{9}. 
We thank Prof. D'Agostini for correspondence.}.

\begin{equation}
P(\lambda_s|\hat x) = \displaystyle
\frac{(\lambda_s + \lambda_b)^{\hat x} }
{\hat x! \displaystyle \sum_{i=0}^{\hat x}{\lambda_b^i \over {i!}}}
\displaystyle e^{-\lambda_s}.
\end{equation}

The numerical results for the
confidence intervals and for comparison the results of paper~\cite{1}
are presented in Table~1 and Table~2.

It should be noted that 
in our approach the dependence of the width of 
confidence intervals for parameter $\lambda_s$ 
on the value of $\lambda_b$ in the case $\hat x = 0$ is absent. 
For $\hat x = 0$ the method proposed in ref.~\cite{10} also gives 
a 90\% upper limit independent of $\lambda_b$. This dependence
is absent also in Bayesian approach~\cite{8,11}.

\section{Conclusion}

The results of construction of frequentist confidence intervals 
for the parameter $\lambda_s$ of Poisson distribution
for the signal in the presence of background with known value
of parameter $\lambda_b$ are presented. It is shown that
the described procedure has both Bayesian and frequentist interpretations.

\begin{flushleft}
{\large \bf Acknowledgments}
\end{flushleft}

We are grateful to V.A.~Matveev, V.F.~Obraztsov and Fred James
for the interest to this work and for valuable comments. 
We wish to thank S.S.~Bityukov, A.V.~Dorokhov, V.A.~Litvine and 
V.N.~Susoikin for useful
discussions.  This work has been supported by RFFI grant 99-02-16956 
and grant INTAS-CERN 377.

\begin{figure}[h]

\centerline{\epsfig{file=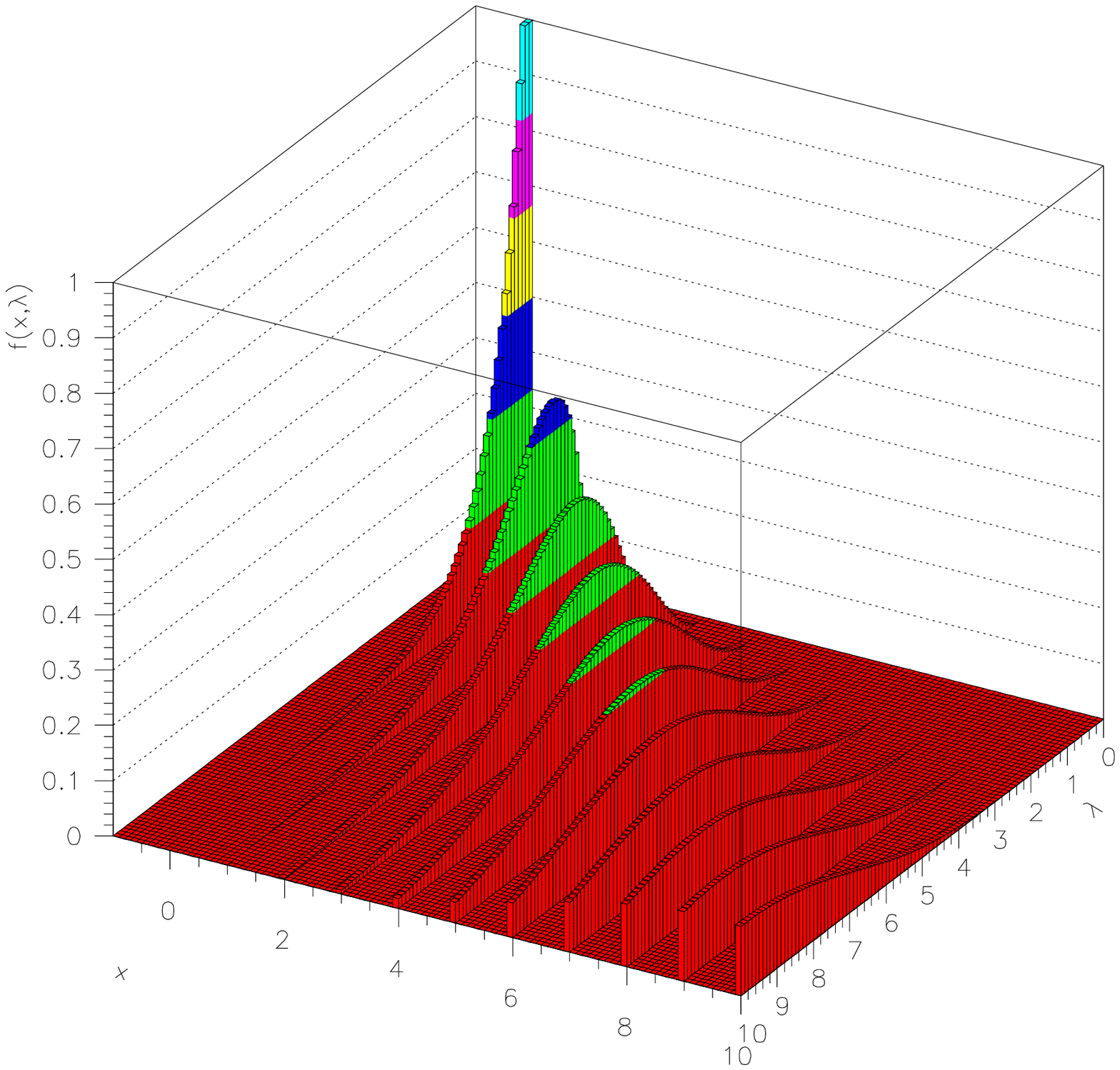,width=14cm}}


\caption{\small The behaviour of the probability density of true value of 
parameter $\lambda$ for Poisson distribution in case of $x$ observed 
events versus $\lambda$ and $x$. Here 
$f(x,\lambda) =\displaystyle \frac{\lambda^x}{x!} e^{-\lambda}$ 
is both a Poisson distribution with parameter $\lambda$ along the axis $x$
and a Gamma distribution with a shape parameter $x+1$ and a scale 
parameter 1 along the axis $\lambda$.
}
\label{fig.1}

\end{figure}

\begin{figure}[h]

\centerline{\epsfig{file=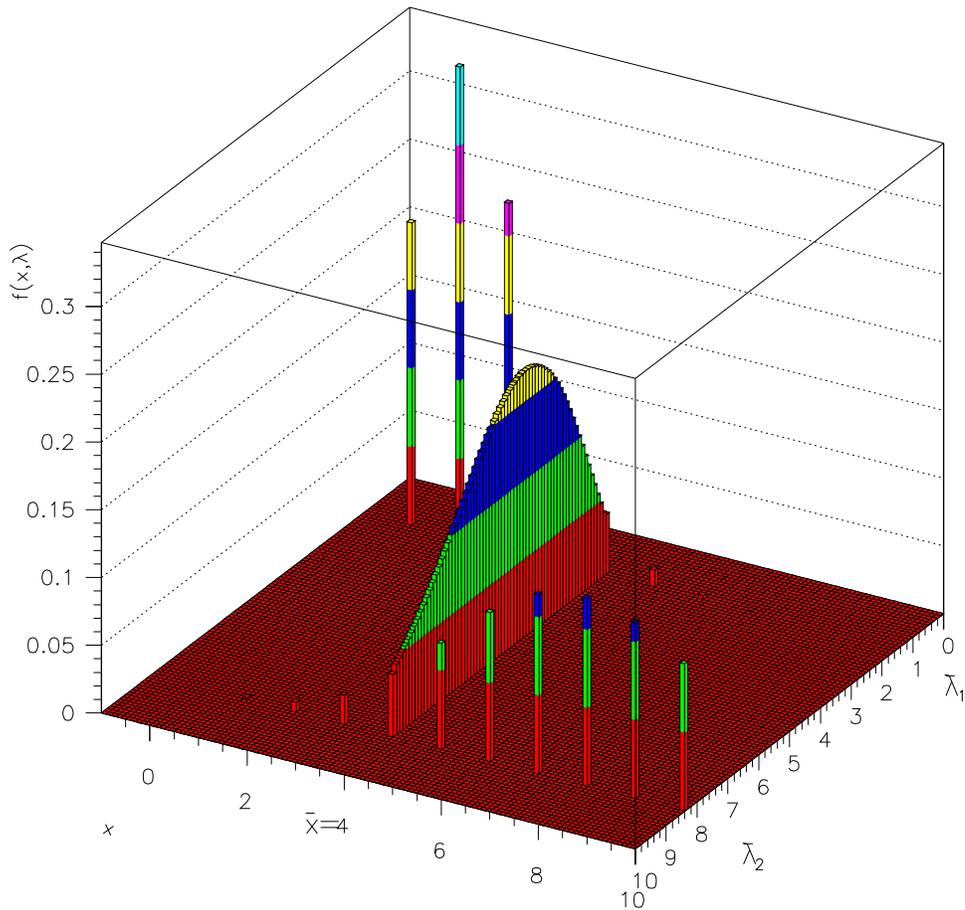,width=14cm}}


\caption{\small The Poisson distributions $f(x,\lambda)$ 
for $\lambda$'s determined by the confidence limits $\hat \lambda_1 = 1.51$ 
and $\hat \lambda_2 = 8.36$ in case of observed number of events 
$\hat x = 4$ are shown. The probability density of Gamma distribution 
with scale parameter $a=1$ and shape parameter  $x=\hat x=4$ is shown 
inside this confidence interval. 
}
\label{fig.2}

\end{figure}

\squeezetable

\widetext

\begin{table}
\caption{90\% C.L. intervals for the Poisson signal mean $\lambda_s$, 
for total events observed $\hat x$, for known mean background $\lambda_b$
ranging from 0 to 4. A comparison between results of ref.[1] and results  
from present note.}
\label{tab.1}
\bigskip


\begin{tabular}{r|cccccccccc} \hline
~$\hat x\backslash\lambda_b$~  & 
    0.0 ref.[1] &  0.0        & 1.0 ref.[1] &    1.0      & 2.0 ref.[1] & 
          2.0      & 3.0 ref.[1] &    3.0      & 4.0 ref.[1] &    4.0     \\ 
\hline
0 &  0.00, 2.44 &  0.00, 2.30 &  0.00, 1.61 &  0.00, 2.30 &  0.00, 1.26 &  
       0.00, 2.30 &  0.00, 1.08 &  0.00, 2.30 &  0.00, 1.01 &  0.00, 2.30 \\
1 &  0.11, 4.36 &  0.09, 3.93 &  0.00, 3.36 &  0.00, 3.27 &  0.00, 2.53 &  
       0.00, 3.00 &  0.00, 1.88 &  0.00, 2.84 &  0.00, 1.39 &  0.00, 2.74 \\
2 &  0.53, 5.91 &  0.44, 5.48 &  0.00, 4.91 &  0.00, 4.44 &  0.00, 3.91 &  
       0.00, 3.88 &  0.00, 3.04 &  0.00, 3.53 &  0.00, 2.33 &  0.00, 3.29 \\
3 &  1.10, 7.42 &  0.93, 6.94 &  0.10, 6.42 &  0.00, 5.71 &  0.00, 5.42 & 
       0.00, 4.93 &  0.00, 4.42 &  0.00, 4.36 &  0.00, 3.53 &  0.00, 3.97 \\
4 &  1.47, 8.60 &  1.51, 8.36 &  0.74, 7.60 &  0.51, 7.29 &  0.00, 6.60 &  
       0.00, 6.09 &  0.00, 5.60 &  0.00, 5.35 &  0.00, 4.60 &  0.00, 4.78 \\
5 &  1.84, 9.99 &  2.12, 9.71 &  1.25, 8.99 &  1.15, 8.73 &  0.43, 7.99 &  
       0.20, 7.47 &  0.00, 6.99 &  0.00, 6.44 &  0.00, 5.99 &  0.00, 5.72 \\
6 &  2.21,11.47 &  2.78,11.05 &  1.61,10.47 &  1.79,10.07 &  1.08, 9.47 &  
       0.83, 9.01 &  0.15, 8.47 &  0.00, 7.60 &  0.00, 7.47 &  0.00, 6.76 \\
7 &  3.56,12.53 &  3.47,12.38 &  2.56,11.53 &  2.47,11.38 &  1.59,10.53 &  
       1.49,10.37 &  0.89, 9.53 &  0.57, 9.20 &  0.00, 8.53 &  0.00, 7.88 \\
8 &  3.96,13.99 &  4.16,13.65 &  2.96,12.99 &  3.18,12.68 &  2.14,11.99 &  
       2.20,11.69 &  1.51,10.99 &  1.21,10.60 &  0.66, 9.99 &  0.34, 9.33 \\
9 &  4.36,15.30 &  4.91,14.95 &  3.36,14.30 &  3.91,13.96 &  2.53,13.30 &  
       2.90,12.94 &  1.88,12.30 &  1.92,11.94 &  1.33,11.30 &  0.97,10.81 \\
10 & 5.50,16.50 &  5.64,16.21 &  4.50,15.50 &  4.66,15.22 &  3.50,14.50 &  
       3.66,14.22 &  2.63,13.50 &  2.64,13.21 &  1.94,12.50 &  1.67,12.16 \\
20 & 13.55,28.52& 13.50,28.33 & 12.55,27.52 & 12.53,27.34 & 11.55,26.52 & 
       11.53,26.34 & 10.55,25.52 & 10.53,25.34 &  9.55,24.52 &  9.53,24.34 \\
\end{tabular}
\end{table}

\begin{table}
\caption{90\% C.L. intervals for the Poisson signal mean $\lambda_s$, 
for total events observed $\hat x$, for known mean background $\lambda_b$
ranging from 6 to 15. A comparison between results of ref.[1] and results  
from present note.}
\label{tab.2}
\bigskip
\begin{tabular}{r|cccccccccc} \hline
~$\hat x\backslash\lambda_b$~  & 
    6.0 ref.[1] &    6.0      & 8.0 ref.[1] &    8.0      & 10.0 ref.[1] &   
            10.0  & 12.0 ref.[1]&   12.0      & 15.0 ref.[1]&   15.0      \\ 
\hline
0 & 0.00, 0.97  & 0.00, 2.30  & 0.00, 0.94  & 0.00, 2.30  & 0.00, 0.93   &
      0.00, 2.30  & 0.00, 0.92  & 0.00, 2.30  & 0.00, 0.92  &  0.00, 2.30  \\
1 & 0.00, 1.14  & 0.00, 2.63  & 0.00, 1.07  & 0.00, 2.56  & 0.00, 1.03   & 
      0.00, 2.51  & 0.00, 1.00  & 0.00, 2.48  & 0.00, 0.98  &  0.00, 2.45  \\
2 &  0.00, 1.57 & 0.00, 3.01  & 0.00, 1.27  & 0.00, 2.85  & 0.00, 1.15   & 
       0.00, 2.75 & 0.00, 1.09  & 0.00, 2.68  & 0.00, 1.05  &  0.00, 2.61  \\
3 &  0.00, 2.14 &  0.00, 3.48 &  0.00, 1.49 &  0.00, 3.20 & 0.00, 1.29   &
       0.00, 3.02 & 0.00, 1.21  & 0.00, 2.91  & 0.00, 1.14  &  0.00, 2.78  \\
4 &  0.00, 2.83 & 0.00, 4.04  &  0.00, 1.98 &  0.00, 3.61 & 0.00, 1.57   &
       0.00, 3.34 & 0.00, 1.37  & 0.00, 3.16  & 0.00, 1.24  &  0.00, 2.98  \\
5 &  0.00, 4.07 &  0.00, 4.71 &  0.00, 2.60 &  0.00, 4.10 & 0.00, 1.85   &
       0.00, 3.72 &  0.00, 1.58 &  0.00, 3.46 & 0.00, 1.32  &  0.00, 3.20 \\
6 &  0.00, 5.47 &  0.00, 5.49 &  0.00, 3.73 &  0.00, 4.67 & 0.00, 2.40   &
       0.00, 4.15 &  0.00, 1.86 &  0.00, 3.80 &  0.00, 1.47 & 0.00, 3.46  \\
7 &  0.00, 6.53 &  0.00, 6.38 &  0.00, 4.58 &  0.00, 5.34 &  0.00, 3.26  &
       0.00, 4.65 &  0.00, 2.23 &  0.00, 4.19 &  0.00, 1.69 &  0.00, 3.74 \\
8 &  0.00, 7.99 &  0.00, 7.35 &  0.00, 5.99 &  0.00, 6.10 &  0.00, 4.22  &
       0.00, 5.23 &  0.00, 2.83 &  0.00, 4.64 &  0.00, 1.95 &  0.00, 4.06 \\
9 &  0.00, 9.30 &  0.00, 8.41 &  0.00, 7.30 &  0.00, 6.95 &  0.00, 5.30  &
       0.00, 5.89 &  0.00, 3.93 &  0.00, 5.15 &  0.00, 2.45 &  0.00, 4.42 \\
10 &  0.22,10.50 &  0.02, 9.53 &  0.00, 8.50 &  0.00, 7.88 &  0.00, 6.50 &
       0.00, 6.63 &  0.00, 4.71 &  0.00, 5.73 &  0.00, 3.00 &  0.00, 4.83 \\
20 &  7.55,22.52 &  7.53,22.34 &  5.55,20.52 &  5.53,20.34 &  3.55,18.52 &
       3.55,18.30 &  2.23,16.52 &  1.70,16.08 &  0.00,13.52 &  0.00,12.31 \\
\end{tabular}
\end{table}

\narrowtext

\end{document}